\documentclass[10pt,final,journal,twocolumn]{IEEEtran}
\ifCLASSINFOpdf

\else
\usepackage{algorithm}
\usepackage{booktabs}

\fi
\usepackage{cite}
\usepackage[cmex10]{amsmath}
\usepackage{amsthm}
\usepackage{stfloats}
\usepackage{multicol,multienum}
\usepackage{multirow}

\usepackage{amssymb}
\usepackage{url}
\usepackage{amsmath}
\usepackage{xcolor}
\usepackage[dvips]{graphicx}
\usepackage{subfigure}
\usepackage{caption}
\usepackage{amsmath}
\usepackage{amsfonts,amssymb}
\usepackage{bbm}
\usepackage[T1]{fontenc}
\usepackage{algorithm}
\usepackage{algpseudocode}
\usepackage{ulem}

\usepackage[linesnumbered, ruled]{}
\makeatletter
\renewcommand{\maketag@@@}[1]{\hbox{\m@th\normalsize\normalfont#1}}%
\makeatother

\begin{document}
\hyphenation{op-tical net-works semi-conduc-tor}
\title{\Huge Secure Wireless Communication via Movable-Antenna Array }
\author{Guojie Hu, Qingqing Wu,~\textit{Senior Member}, \textit{IEEE}, Kui Xu,~\textit{Member}, \textit{IEEE}, Jiangbo Si,~\textit{Senior Member}, \textit{IEEE}, and Naofal Al-Dhahir,~\textit{Fellow}, \textit{IEEE}
\thanks{
%
Guojie Hu is with the College of Communication Engineering, Rocket Force University of Engineering, Xi'an 710025, China (email: lgdxhgj@sina.com). Qingqing Wu is with the Department
of Electronic Engineering, Shanghai Jiao Tong University, Shanghai 200240, China. Kui Xu is with the College of Communications Engineering, the Army of Engineering University, Nanjing 210007, China. Jiangbo Si is with the Integrated Service Networks Lab of Xidian University, Xi'an 710100, China. Naofal Al-Dhahir is with the Department of Electrical and Computer Engineering, The University of Texas at Dallas, Richardson, TX 75080 USA.
}
}
\IEEEpeerreviewmaketitle
\maketitle
\begin{abstract}
Movable antenna (MA) array is a novel technology recently developed where positions of transmit/receive antennas can be flexibly adjusted in the specified region to reconfigure the wireless channel and achieve a higher capacity. In this letter, we, for the first time, investigate the MA array-assisted physical-layer security where the confidential information is transmitted from a MA array-enabled Alice to a single-antenna Bob, in the presence of multiple single-antenna and colluding eavesdroppers. We aim to maximize the achievable secrecy rate by jointly designing the transmit beamforming and positions of all antennas at Alice subject to the transmit power budget and specified regions for positions of all transmit antennas. The resulting problem is highly non-convex, for which the projected gradient ascent (PGA) and the alternating optimization methods are utilized to obtain a high-quality suboptimal solution. Simulation results demonstrate that since the additional spatial degree of freedom (DoF) can be fully exploited, the MA array significantly enhances the secrecy rate compared to the conventional fixed-position antenna (FPA) array.
\end{abstract}


\IEEEpeerreviewmaketitle
\vspace{-5pt}
\section{Introduction}
In conventional physical-layer security (PLS) systems, beamforming is one of the key techniques to strengthen/degrade the signal reception quality at desired/undesired directions to enhance the communication security [1]. Existing beamforming techniques rely on the fixed-position antenna (FPA) array, where all antennas are deployed at fixed positions. This brings a fundamental problem, i.e., the steering vector of the FPA array is static corresponding to a fixed steering angle, which will weaken the secure beamforming gain because of the inherent spatial correlation over different (desired and undesired) steering angles [2].

To overcome this limitation, movable antenna (MA) technology is recently proposed [3]$-$[4]. In addition to beamforming, an additional spatial degree of freedom (DoF) is exploited by the MA array, i.e., positions of all antennas can be flexibly adjusted in a specified region to vary the steering vectors corresponding to different angles, i.e., reconfigure the wireless channels, and achieve higher communication capacity. Previous works have demonstrated the great potential of MA array-enabled communications [3]$-$[13]. The promising applications, hardware architecture, channel characterization and performance advantages of the MA array are introduced in [3]$-$[4]. Compressed sensing based channel estimation for MA array-enabled communications is designed in [5]. The joint designs of transmit covariance and antenna movement for a multiple-input multiple-output (MIMO) system are provided in [6]$-$[7]. The channel model and performance analysis of outage probability for MA array-enabled systems are characterized in [8]. Exploiting the MA array for concurrently enhancing/nulling the signal power at desired/undesired directions is studied in [9]. The extensions to MA array-enabled multi-user communications are investigated in [10]$-$[13].

Different from all previous works, this letter considers the application of the MA array to enhance secure communication from a MA array-enabled Alice to a single-antenna Bob in the presence of multiple single-antenna eavesdroppers. Our objective is to maximize the achievable secrecy rate, by jointly optimizing the transmit beamforming and positions of all antennas at Alice. The projected gradient ascent (PGA) method and the alternating optimization algorithm are developed to solve the formulated highly non-convex problem. It is shown by simulations that the MA array greatly improves the secrecy rate compared to the conventional FPA array.

 \newcounter{mytempeqncnt}
 \vspace{-5pt}
\section{System Model and Problem Formulation}
As illustrated in Fig. 1, we consider a secrecy communication system where Alice intends to transmit confidential information to a legitimate destination (Bob) in the presence of $M$ eavesdroppers (Eves). Bob and each of $M$ Eves are equipped with a single and fixed-position antenna, while Alice is equipped with a linear MA array of size $N$. The position of the $n$-th antenna at Alice is denoted by $x_n$, $1 \le n \le N$, and positions of $N$ antennas can be expressed compactly as ${\bf{x}} = {\left[ {{x_1},{x_2},...,{x_N}} \right]^T} \in {{\mathbb{R}}^{N \times 1}}$ where ${( \cdot )^T}$ is the transpose operation. Therefore, given ${\bf{x}}$ and the steering angle with respect to (w.r.t.) the linear MA array as $\theta $, the corresponding steering vector of the MA array is given by
{\setlength\abovedisplayskip{3.5pt}
\setlength\belowdisplayskip{3.5pt}
\begin{equation}
\begin{split}{}
{\bf{a}}\left( {{\bf{x}},\theta } \right) = {\left[ {{e^{j\frac{{2\pi }}{\lambda }{x_1}\cos \theta }},{e^{j\frac{{2\pi }}{\lambda }{x_2}\cos \theta }},...,{e^{j\frac{{2\pi }}{\lambda }{x_N}\cos \theta }}} \right]^T},
\end{split}
\end{equation}
where $\lambda $ is the wavelength. Then, suppose that ${\bf{w}} \in {{\mathbb{C}}^{N \times 1}}$ is the digital transmit beamforming at Alice for the confidential information, the beam gain of the MA array at the angle $\theta $ is derived as
\begin{equation}
{G_{{\bf{x}},{\bf{w}}}}(\theta ) = {\left| {{{\bf{a}}^H}\left( {{\bf{x}},\theta } \right){\bf{w}}} \right|^2},\theta  \in [0,\pi ),
\end{equation}
where ${( \cdot )^H}$ is the conjugate transpose operation.

 \begin{figure}
\centering
\includegraphics[width=7cm]{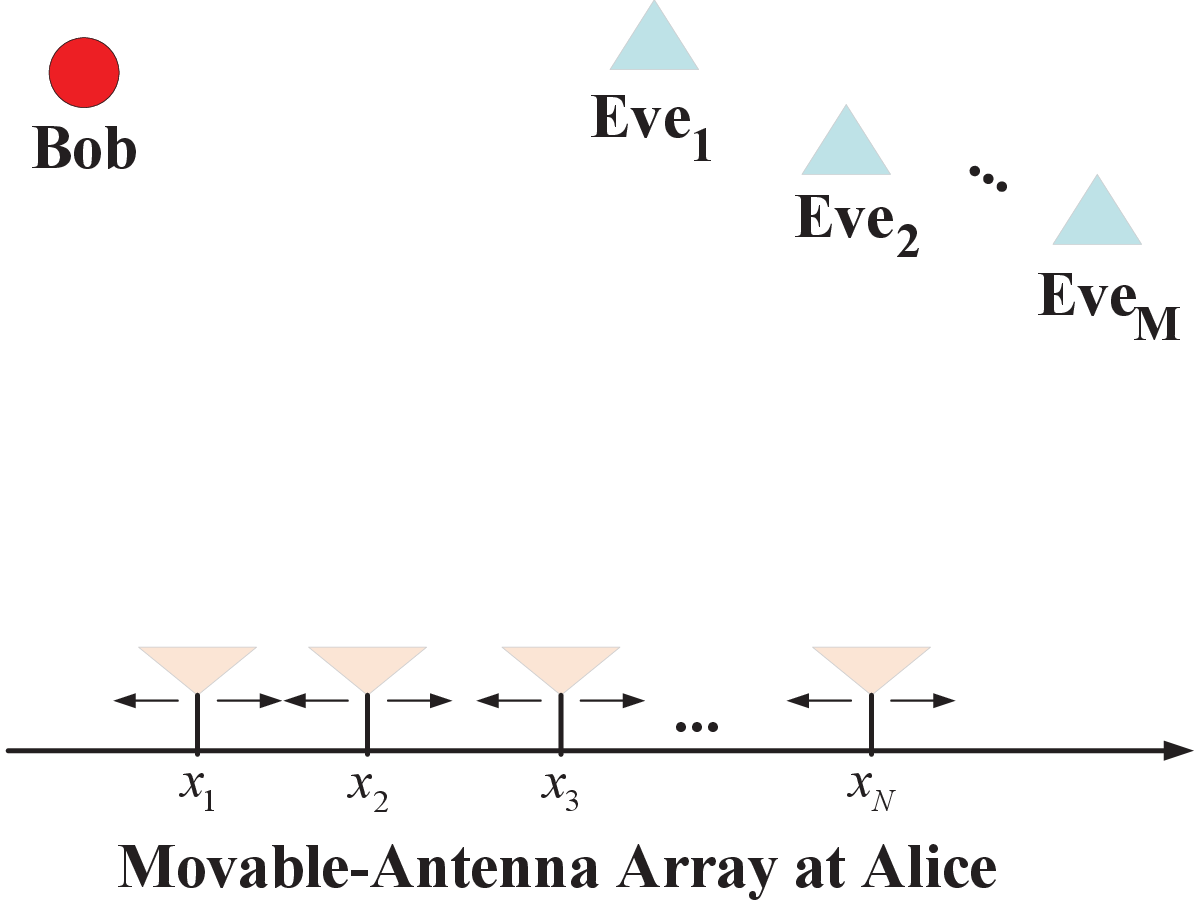}
\captionsetup{font=small}
\caption{Illustration of the system model.} \label{fig:Fig1}
\vspace{-15pt}
\end{figure}

Based on (2) and considering the worst (colluding) case where $M$ Eves aim to cooperatively process their received confidential information, the achievable secrecy rate in bits/second/Hz (bps/Hz) is given by [14]\footnotemark \footnotetext{Here, we omit the large-scale fading power for the main channel and $M$ eavesdropping channels, as such power is normalized over the receiver noise power.}
\begin{equation}
{R_{\sec }}({\bf{x}},{\bf{w}}) = {\left[ \begin{array}{l}
{\log _2}\left( {1 + \frac{{{{\left| {{{\bf{a}}^H}\left( {{\bf{x}},{\theta _0}} \right){\bf{w}}} \right|}^2}}}{{{\sigma ^2}}}} \right)\\
 - {\log _2}\left( {1 + \frac{{\sum\nolimits_{i = 1}^M {{{\left| {{{\bf{a}}^H}\left( {{\bf{x}},{\theta _i}} \right){\bf{w}}} \right|}^2}} }}{{{\sigma ^2}}}} \right)
\end{array} \right]^ + },
\end{equation}
where ${\left[ r \right]^ + } = \max (r,0)$, ${{\theta _0}}$ is the steering angle to where Bob is located, ${{\theta _i}}$ is the steering angle to where the $i$-th Eve is located and ${{\sigma ^2}}$ is the receiver noise power.

Our goal is to maximize ${R_{\sec }}({\bf{x}},{\bf{w}})$, by jointly optimizing the positions of the MA array ${\bf{x}}$ and the transmit beamforming ${\bf{w}}$ at Alice. Hence, the optimization problem is formulated as
 \begin{align}
&({\rm{P1}}):{\rm{  }}\mathop {\max }\limits_{{{\bf{x}},{\bf{w}}}} \ {R_{\sec }}({\bf{x}},{\bf{w}})\tag{${\rm{4a}}$}\\
{\rm{              }}&\ {\rm{s.t.}} \quad \left| {{x_s} - {x_c}} \right| \ge {d_{\min }},1 \le s \ne c \le N,\tag{${\rm{4b}}$}\\
&\quad \quad \ \ \left\{ {{x_s}} \right\}_{s = 1}^N \in [0,L],\tag{${\rm{4c}}$}\\
 & \quad \ \ \quad \left\| {\bf{w}} \right\|_2^2 = {P_A},\tag{${\rm{4d}}$}
 \end{align}
where ${d_{\min }}$ in (4b) is the minimum distance between any two MAs for avoiding the coupling effect, (4c) provides the range of $\left\{ {{x_s}} \right\}_{s = 1}^N$ and $P_A$ in (4d) is the power budget of Alice. To make (4b) always hold, it is clear to determine that $L$ in (4c) should be no smaller than $(N - 1){d_{\min }}$.


\textbf{Remark 1}: Since the MA array allows the flexible adjustment for the positions of all antennas, predictably it can reap better spatial diversity and multiplexing performance compared to the conventional FPA array (which is a special case of the considered MA array) and thus achieve a higher secrecy rate. However, the objective of (P1) is non-concave w.r.t. ${\bf{x}}$ or ${\bf{w}}$. Moreover, ${\bf{x}}$ and ${\bf{w}}$ are coupled with each other in the objective. These two aspects result in the high non-convexity of (P1). Consequently, the alternating optimization algorithm will be exploited to iteratively solve ${\bf{x}}$ or ${\bf{w}}$ with the other variable being fixed.

\section{Alternating Optimization for (P1)}
\subsection{Optimizing ${\bf{w}}$ Given ${\bf{x}}$}
Let us first define
{\setlength\abovedisplayskip{4pt}
\setlength\belowdisplayskip{4pt}
\begin{equation}
\setcounter{equation}{5}
\begin{split}{}
{\bf{A}} =& \frac{1}{{{\sigma ^2}}}{\bf{a}}\left( {{\bf{x}},{\theta _0}} \right){{\bf{a}}^H}\left( {{\bf{x}},{\theta _0}} \right),\\
{\bf{B}} =& \frac{1}{{{\sigma ^2}}}\sum\nolimits_{i = 1}^M {{\bf{a}}\left( {{\bf{x}},{\theta _i}} \right){{\bf{a}}^H}\left( {{\bf{x}},{\theta _i}} \right)},
\end{split}
\end{equation}
based on which the problem of optimizing ${\bf{w}}$ given ${\bf{x}}$ is expressed as\footnotemark \footnotetext{The operator ${\left[  \cdot  \right]^ + }$ can be omitted without loss of optimality. The same operation is applied to the following problem (P3).}
 \begin{align}
&({\rm{P2}}):{\rm{  }}\mathop {\max }\limits_{{{\bf{w}}}} \ \frac{{1 + {{\bf{w}}^H}{\bf{Aw}}}}{{1 + {{\bf{w}}^H}{\bf{Bw}}}} \tag{${\rm{6a}}$}\\
{\rm{              }}&\ {\rm{s.t.}} \quad \left\| {\bf{w}} \right\|_2^2 = {P_A}.\tag{${\rm{6b}}$}
 \end{align}

The optimal solution to (P2) is well-established as [15]
\begin{equation}
\setcounter{equation}{7}
{\bf{w}} = \sqrt {{P_A}} {{\bf{o}}_{\max }},
\end{equation}
where ${{\bf{o}}_{\max }}$ is the normalized eigenvector corresponding to the largest eigenvalue of the matrix ${\left( {{\bf{B}} + \frac{1}{{{P_A}}}{{\bf{I}}_N}} \right)^{ - 1}}\left( {{\bf{A}} + \frac{1}{{{P_A}}}{{\bf{I}}_N}} \right)$, ${{{\bf{I}}_N}}$ is the $N \times N$ identity matrix and ${\left(  \cdot  \right)^{ - 1}}$ is the inverse operation.

 \begin{figure*}[ht]
  \vspace{-5pt}
\setcounter{mytempeqncnt}{\value{equation}}
\setcounter{equation}{18}
\begin{equation}
\begin{split}{}
{\nabla _{{{\bf{x}}^t}}}\Psi \left( {\left\{ {{{\bf{g}}_i},{{\bf{q}}_i}} \right\}_{i = 0}^M} \right) = \ln 2\left( {\frac{{ - \frac{{{{\bf{W}}_0}(2{\bf{C}}{{\bf{g}}_0} + 2{\bf{D}}{{\bf{q}}_0}) + {{\bf{S}}_0}(2{\bf{C}}{{\bf{q}}_0} - 2{\bf{D}}{{\bf{g}}_0})}}{{{\sigma ^2}}}}}{{1 + f\left( {{{\bf{g}}_0},{{\bf{q}}_0}} \right)/{\sigma ^2}}} + \frac{{\sum\nolimits_{i = 1}^M {\frac{{{{\bf{W}}_i}(2{\bf{C}}{{\bf{g}}_i} + 2{\bf{D}}{{\bf{q}}_i}) + {{\bf{S}}_i}(2{\bf{C}}{{\bf{q}}_i} - 2{\bf{D}}{{\bf{g}}_i})}}{{{\sigma ^2}}}} }}{{1 + \sum\nolimits_{i = 1}^M {f\left( {{{\bf{g}}_i},{{\bf{q}}_i}} \right)} /{\sigma ^2}}}} \right).
\end{split}
\end{equation}
\setcounter{equation}{\value{mytempeqncnt}}
\hrulefill
\vspace{-5pt}
\end{figure*}

\subsection{Optimizing ${\bf{x}}$ Given ${\bf{w}}$}
Let us first define
\begin{equation}
\begin{split}{}
{{\bf{g}}_i} =& {\left[ {{g_{1,i}},{g_{2,i}},...,{g_{N,i}}} \right]^T},i = 0,1,...,M,\\
{{\bf{q}}_i} =& {\left[ {{q_{1,i}},{q_{2,i}},...,{q_{N,i}}} \right]^T},i = 0,1,...,M,
\end{split}
\end{equation}
with
\begin{equation}
\begin{split}{}
{g_{n,i}} =& \cos \left( {\frac{{2\pi }}{\lambda }{x_n}\cos {\theta _i}} \right),{q_{n,i}} = \sin \left( {\frac{{2\pi }}{\lambda }{x_n}\cos {\theta _i}} \right).
\end{split}
\end{equation}

Further, denote ${\bf{w}} = {\bf{u}} + j{\bf{z}}$, ${\bf{C}}  = {\bf{u}}{{\bf{u}}^T} + {\bf{z}}{{\bf{z}}^T}$ and ${\bf{D}} = {\bf{u}}{{\bf{z}}^T} - {\bf{z}}{{\bf{u}}^T}$. Armed with these, ${\left| {{{\bf{a}}^H}({\bf{x}},{\theta _i}){\bf{w}}} \right|^2}$, $i = 0,1,...,M$, can be derived as
\begin{equation}
\begin{split}{}
{\left| {{{\bf{a}}^H}({\bf{x}},{\theta _i}){\bf{w}}} \right|^2} =& {{\bf{g}}_i}^T{\bf{C}}{{\bf{g}}_i} + {{\bf{q}}_i}^T{\bf{C}}{{\bf{q}}_i} + 2{{\bf{g}}_i}^T{\bf{D}}{{\bf{q}}_i}\\
 \buildrel \Delta \over =& f\left( {{{\bf{g}}_i},{{\bf{q}}_i}} \right),
\end{split}
\end{equation}
based on which the problem of optimizing ${\bf{x}}$ given ${\bf{w}}$ is expressed as
 \begin{align}
&({\rm{P3}}):{\rm{  }}\mathop {\max }\limits_{{{\bf{x}}}} \ {\log _2}\left( {1 + \frac{{f\left( {{{\bf{g}}_0},{{\bf{q}}_0}} \right)}}{{{\sigma ^2}}}} \right) \tag{${\rm{11a}}$}\\
&\quad \quad \quad- {\log _2}\left( {1 + \frac{{\sum\nolimits_{i = 1}^M {f\left( {{{\bf{g}}_i},{{\bf{q}}_i}} \right)} }}{{{\sigma ^2}}}} \right) \buildrel \Delta \over = \Psi \left( {\left\{ {{{\bf{g}}_i},{{\bf{q}}_i}} \right\}_{i = 0}^M} \right) \nonumber\\
{\rm{              }}&\ {\rm{s.t.}} \quad (4{\rm{b}}), (4{\rm{c}})\tag{${\rm{11b}}$}.
 \end{align}

Problem (P3) is still highly non-convex due to the complex objective. To tackle this issue, the projected gradient ascent (PGA) method [16] is exploited to find a locally optimal solution to (P3). Specifically, using the PGA, the update rule for ${\bf{x}}$ is given by:
\begin{equation}
\setcounter{equation}{12}
\begin{split}{}
{{\bf{x}}^{t + 1}} =& {{\bf{x}}^t} + \delta {\nabla _{{{\bf{x}}^t}}}\Psi \left( {\left\{ {{{\bf{g}}_i},{{\bf{q}}_i}} \right\}_{i = 0}^M} \right),\\
{{\bf{x}}^{t + 1}} =& {\cal B}\left\{ {{{\bf{x}}^{t + 1}},{d_{\min }},L} \right\},
\end{split}
\end{equation}
where ${{{\bf{x}}^{t+1}}}$ in the first equation of (12) is the original updated ${\bf{x}}$ in the $t+1$-th iteration, and ${{{\bf{x}}^{t+1}}}$ in the second equation of (12) is the additional update (if necessary) via the projection function ${\cal B}\left\{  \cdot  \right\}$ as explained later, which ensures that the solutions for MA positions in each iteration always satisfy the constraint in (11b). In addition, ${{\nabla _{{{\bf{x}}^t}}}\Psi \left( {\left\{ {{{\bf{g}}_i},{{\bf{q}}_i}} \right\}_{i = 0}^M} \right)}$ denotes the gradient of ${\Psi \left( {\left\{ {{{\bf{g}}_i},{{\bf{q}}_i}} \right\}_{i = 0}^M} \right)}$ w.r.t. ${{{\bf{x}}^t}}$, and $\delta $ is the step size for gradient ascent.

\subsubsection{Computing ${{\nabla _{{{\bf{x}}^t}}}\Psi \left( {\left\{ {{{\bf{g}}_i},{{\bf{q}}_i}} \right\}_{i = 0}^M} \right)}$}
In detail, the gradient of $\Psi \left( {\left\{ {{{\bf{g}}_i},{{\bf{q}}_i}} \right\}_{i = 0}^M} \right)$ w.r.t. ${{{\bf{x}}^t}}$ is derived as
\begin{equation}
\begin{split}{}
&{\nabla _{{{\bf{x}}^t}}}\Psi \left( {\left\{ {{{\bf{g}}_i},{{\bf{q}}_i}} \right\}_{i = 0}^M} \right) = \ln 2\\
 \times & \left( {\frac{{{\nabla _{{{\bf{x}}^t}}}f\left( {{{\bf{g}}_0},{{\bf{q}}_0}} \right)/{\sigma ^2}}}{{1 + f\left( {{{\bf{g}}_0},{{\bf{q}}_0}} \right)/{\sigma ^2}}} - \frac{{\sum\nolimits_{i = 1}^M {{\nabla _{{{\bf{x}}^t}}}f\left( {{{\bf{g}}_i},{{\bf{q}}_i}} \right)/{\sigma ^2}} }}{{1 + \sum\nolimits_{i = 1}^M {f\left( {{{\bf{g}}_i},{{\bf{q}}_i}} \right)} /{\sigma ^2}}}} \right),
\end{split}
\end{equation}
with
\begin{equation}
{\nabla _{{{\bf{x}}^t}}}f\left( {{{\bf{g}}_i},{{\bf{q}}_i}} \right) = {\left[ {\frac{{\partial f\left( {{{\bf{g}}_i},{{\bf{q}}_i}} \right)}}{{\partial x_1^t}},\frac{{\partial f\left( {{{\bf{g}}_i},{{\bf{q}}_i}} \right)}}{{\partial x_2^t}},...,\frac{{\partial f\left( {{{\bf{g}}_i},{{\bf{q}}_i}} \right)}}{{\partial x_N^t}}} \right]^T},
\end{equation}
for any $i = 0,1,...,M$. Further, for any $n = 1,2,...,N$, $\frac{{\partial f\left( {{{\bf{g}}_i},{{\bf{q}}_i}} \right)}}{{\partial x_n^t}}$ can be derived as
\begin{equation}
\begin{split}{}
&\frac{{\partial f\left( {{{\bf{g}}_i},{{\bf{q}}_i}} \right)}}{{\partial x_n^t}} = \frac{{\partial f\left( {{{\bf{g}}_i},{{\bf{q}}_i}} \right)}}{{\partial {g_{n,i}}}}\frac{{\partial {g_{n,i}}}}{{\partial x_n^t}} + \frac{{\partial f\left( {{{\bf{g}}_i},{{\bf{q}}_i}} \right)}}{{\partial {q_{n,i}}}}\frac{{\partial {q_{n,i}}}}{{\partial x_n^t}}\\
 =&  - \frac{{2\pi }}{\lambda }\cos {\theta _i}\sin \left( {\frac{{2\pi }}{\lambda }{x_n}\cos {\theta _i}} \right)\frac{{\partial f\left( {{{\bf{g}}_i},{{\bf{q}}_i}} \right)}}{{\partial {g_{n,i}}}}\\
& + \frac{{2\pi }}{\lambda }\cos {\theta _i}\cos \left( {\frac{{2\pi }}{\lambda }{x_n}\cos {\theta _i}} \right)\frac{{\partial f\left( {{{\bf{g}}_i},{{\bf{q}}_i}} \right)}}{{\partial {q_{n,i}}}},
\end{split}
\end{equation}
based on which we denote
\begin{equation}
\begin{split}{}
{{\bf{W}}_i} =& {\rm{diag}}\left( {\left\{ {\frac{{2\pi }}{\lambda }\cos {\theta _i}\sin \left( {\frac{{2\pi }}{\lambda }{x_n}\cos {\theta _i}} \right)} \right\}_{n = 1}^N} \right),\\
{{\bf{S}}_i} =& {\rm{diag}}\left( {\left\{ {\frac{{2\pi }}{\lambda }\cos {\theta _i}\cos \left( {\frac{{2\pi }}{\lambda }{x_n}\cos {\theta _i}} \right)} \right\}_{n = 1}^N} \right),
\end{split}
\end{equation}
with ${\rm{diag}}( \cdot )$ denoting the diagonal operation. In addition, based on (7), we have
\begin{equation}
\begin{split}{}
&{\left[ {\frac{{\partial f\left( {{{\bf{g}}_i},{{\bf{q}}_i}} \right)}}{{\partial {g_{1,i}}}},\frac{{\partial f\left( {{{\bf{g}}_i},{{\bf{q}}_i}} \right)}}{{\partial {g_{2,i}}}},...,\frac{{\partial f\left( {{{\bf{g}}_i},{{\bf{q}}_i}} \right)}}{{\partial {g_{N,i}}}}} \right]^T}\\
& \buildrel \Delta \over = {\nabla _{{{\bf{g}}_i}}}f\left( {{{\bf{g}}_i},{{\bf{q}}_i}} \right) = 2{\bf{C}}{{\bf{g}}_i} + 2{\bf{D}}{{\bf{q}}_i},\\
&{\left[ {\frac{{\partial f\left( {{{\bf{g}}_i},{{\bf{q}}_i}} \right)}}{{\partial {q_{1,i}}}},\frac{{\partial f\left( {{{\bf{g}}_i},{{\bf{q}}_i}} \right)}}{{\partial {q_{2,i}}}},...,\frac{{\partial f\left( {{{\bf{g}}_i},{{\bf{q}}_i}} \right)}}{{\partial {q_{N,i}}}}} \right]^T}\\
& \buildrel \Delta \over = {\nabla _{{{\bf{q}}_i}}}f\left( {{{\bf{g}}_i},{{\bf{q}}_i}} \right) = 2{\bf{C}}{{\bf{q}}_i} - 2{\bf{D}}{{\bf{g}}_i}.
\end{split}
\end{equation}

From the above analysis, we can finally derive that
\begin{equation}
\begin{split}{}
&{\nabla _{{{\bf{x}}^t}}}f\left( {{{\bf{g}}_i},{{\bf{q}}_i}} \right)\\
 =&  - {{\bf{W}}_i}(2{\bf{C}}{{\bf{g}}_i} + 2{\bf{D}}{{\bf{q}}_i}) + {{\bf{S}}_i}(2{\bf{C}}{{\bf{q}}_i} - 2{\bf{D}}{{\bf{g}}_i}),
\end{split}
\end{equation}
leading to the closed-form expression for ${\nabla _{{{\bf{x}}^t}}}\Psi \left( {\left\{ {{{\bf{g}}_i},{{\bf{q}}_i}} \right\}_{i = 0}^M} \right)$ shown in (19).

\subsubsection{Determining ${\cal B}\left\{ {{{\bf{x}}^{t + 1}},{d_{\min }},L} \right\}$}
Suppose that $0 \le {x_1} < {x_2} < ... < {x_N} \le L$ without loss of generality. Then, it can be derived that
\begin{equation}
\setcounter{equation}{20}
\begin{split}{}
&{x_2} - {x_1} \ge {d_{\min }},...,{x_N} - {x_{N - 1}} \ge {d_{\min }}\\
 \Rightarrow & {x_N} \ge {x_{N - 1}} + {d_{\min }} \ge ...\\
 &\ge {x_n} + (N - n){d_{\min }} \ge ... \ge {x_1} + (N - 1){d_{\min }}\\
\mathop  \Rightarrow \limits^{{x_N} \le L} & {x_n} + (N - n){d_{\min }} \le L,\forall n = 1,...,N,\\
 \Rightarrow & {x_n} \in \left[ {{x_{n - 1}} + {d_{\min }},L - (N - n){d_{\min }}} \right],\forall n = 1,...,N,
\end{split}
\end{equation}
where we additionally denote ${x_0} =  - {d_{\min }}$ for convenience.

 \begin{algorithm}
\caption{Alternating Optimization for Solving (P1)}
  \begin{algorithmic}[1]
\State Initialize the MA array positions at Alice as ${{\bf{x}}^I} = [0,{d_{\min }},...,(N - 1){d_{\min }}]{}^T$.

\State \textbf{Repeat:}

\State \quad Update ${\bf{w}}$ based on (7).

\State \quad Update ${\bf{x}}$ iteratively based on (12) until

\noindent \quad $\Psi \left( {\left\{ {{{\bf{g}}_i},{{\bf{q}}_i}} \right\}_{i = 0}^M} \right)$ converges to a constant.

\State \textbf{Until:} The objective of (P1) converges to a prescribed accuracy.
  \end{algorithmic}
\end{algorithm}

Equation (20) determines the feasible range for each $x_n$, $n = 1,...,N$. Then, in the $t+1$-th iteration, if $x_1^{t + 1}$ obtained via the first equation of (12) is not in the range of $\left[ {0,L - (N - 1){d_{\min }}} \right]$, we should further update it as $x_1^{t + 1} = \max \left( {0,\min \left( {L - (N - 1){d_{\min }},x_1^{t + 1}} \right)} \right)$ according to the rule of the nearest distance [10]. Otherwise, $x_1^{t + 1}$ obtained via the first equation of (12) does not need to be updated via the second equation of (12). The same procedure should be implemented in turn for updating $x_2^{t + 1},...,x_N^{t + 1}$. In summary, the projection function ${\cal B}\left\{ {{{\bf{x}}^{t + 1}},{d_{\min }},L} \right\}$ can be derived as
\begin{equation}
\begin{split}{}
&{\cal B}\left\{ {{{\bf{x}}^{t + 1}},{d_{\min }},L} \right\}:\\
&\left\{ {\begin{array}{*{20}{c}}
{x_1^{t + 1} = \max \left( {0,\min \left( {L - (N - 1){d_{\min }},x_1^{t + 1}} \right)} \right),}\\
{x_2^{t + 1} = \max \left( {x_1^{t + 1} + {d_{\min }},\min \left( {L - (N - 2){d_{\min }},x_2^{t + 1}} \right)} \right),}\\
{...}\\
{x_N^{t + 1} = \max \left( {x_{N - 1}^{t + 1} + {d_{\min }},\min \left( {L,x_N^{t + 1}} \right)} \right).}
\end{array}} \right.
\end{split}
\end{equation}

With ${\nabla _{{{\bf{x}}^t}}}\Psi \left( {\left\{ {{{\bf{g}}_i},{{\bf{q}}_i}} \right\}_{i = 0}^M} \right)$ and ${\cal B}\left\{ {{{\bf{x}}^{t + 1}},{d_{\min }},L} \right\}$ at hand, the PGA method iteratively updates ${{\bf{x}}^{t + 1}}$ based on (12) until the objective of (P3) converges to a constant value. The computational complexity is about ${\cal O}\left( {((M + 1)N + {{(M + 1)}^2})T_{\rm{{inner}}}} \right)$ [16], where $T_{\rm{{inner}}}$ is the number of iterations in the inner layer.
\subsection{Alternating Optimization}
The overall alternating algorithm for solving (P1) is presented in Algorithm 1. Since ${R_{\sec }}({\bf{x}},{\bf{w}})$ is non-decreasing over iterations and has an upper bound, Algorithm 1 is guaranteed to converge. In addition, the computational complexity of solving (P2) is about ${\cal O}\left( {{N^3}} \right)$ [17]. Hence, the overall complexity of solving (P1) is about ${\cal O}\left( {\left( {{N^3} + ((M + 1)N + {{(M + 1)}^2}){T_{{\rm{inner}}}}} \right){T_{{\rm{outer}}}}} \right)$, where $T_{\rm{{outer}}}$ is the number of iterations in the outer layer.

 \begin{figure}
\centering
\includegraphics[width=9cm]{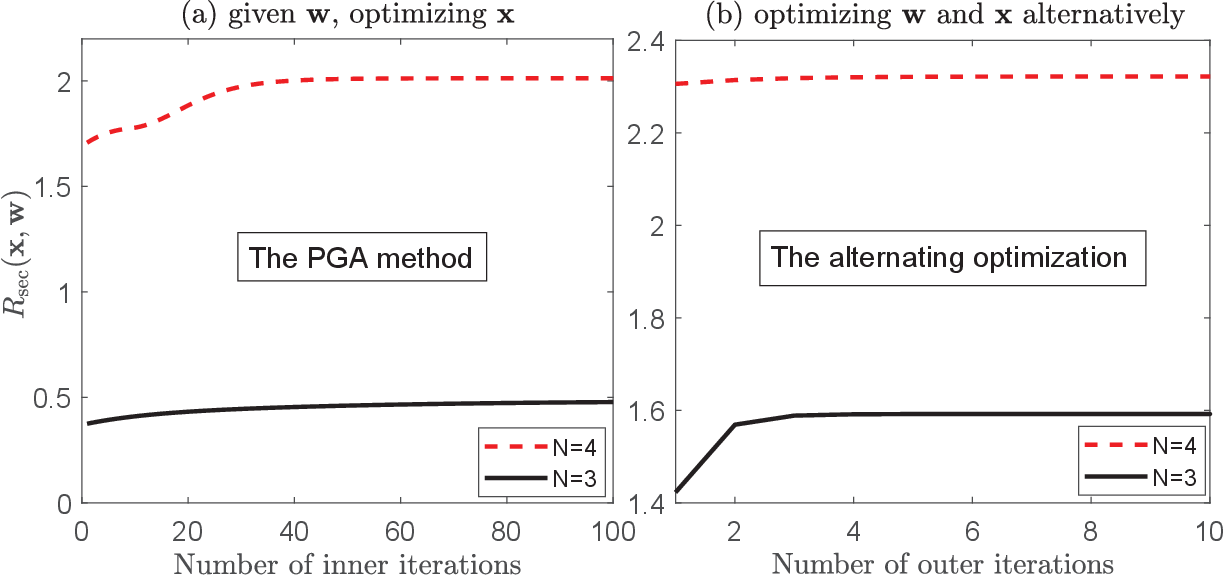}
\captionsetup{font=small}
\caption{Convergence behavior of the PGA method (sub-figure a) and the alternating optimization algorithm (sub-figure b).} \label{fig:Fig1}
\vspace{-10pt}
\end{figure}

\begin{figure}
\centering
\includegraphics[width=8cm]{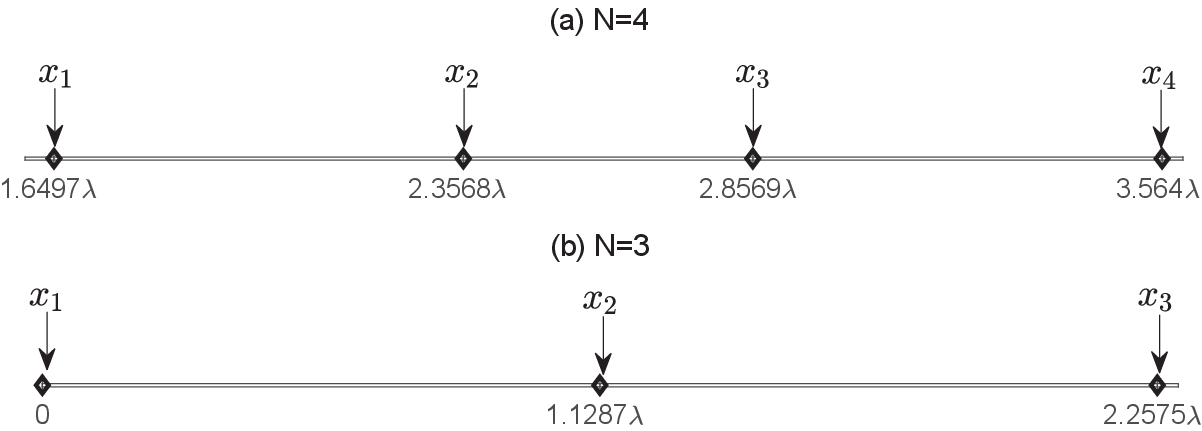}
\captionsetup{font=small}
\caption{Optimized positions of the MA array.} \label{fig:Fig1}
\vspace{-20pt}
\end{figure}

 \begin{figure}
 \vspace{-10pt}
\centering
\includegraphics[width=8cm]{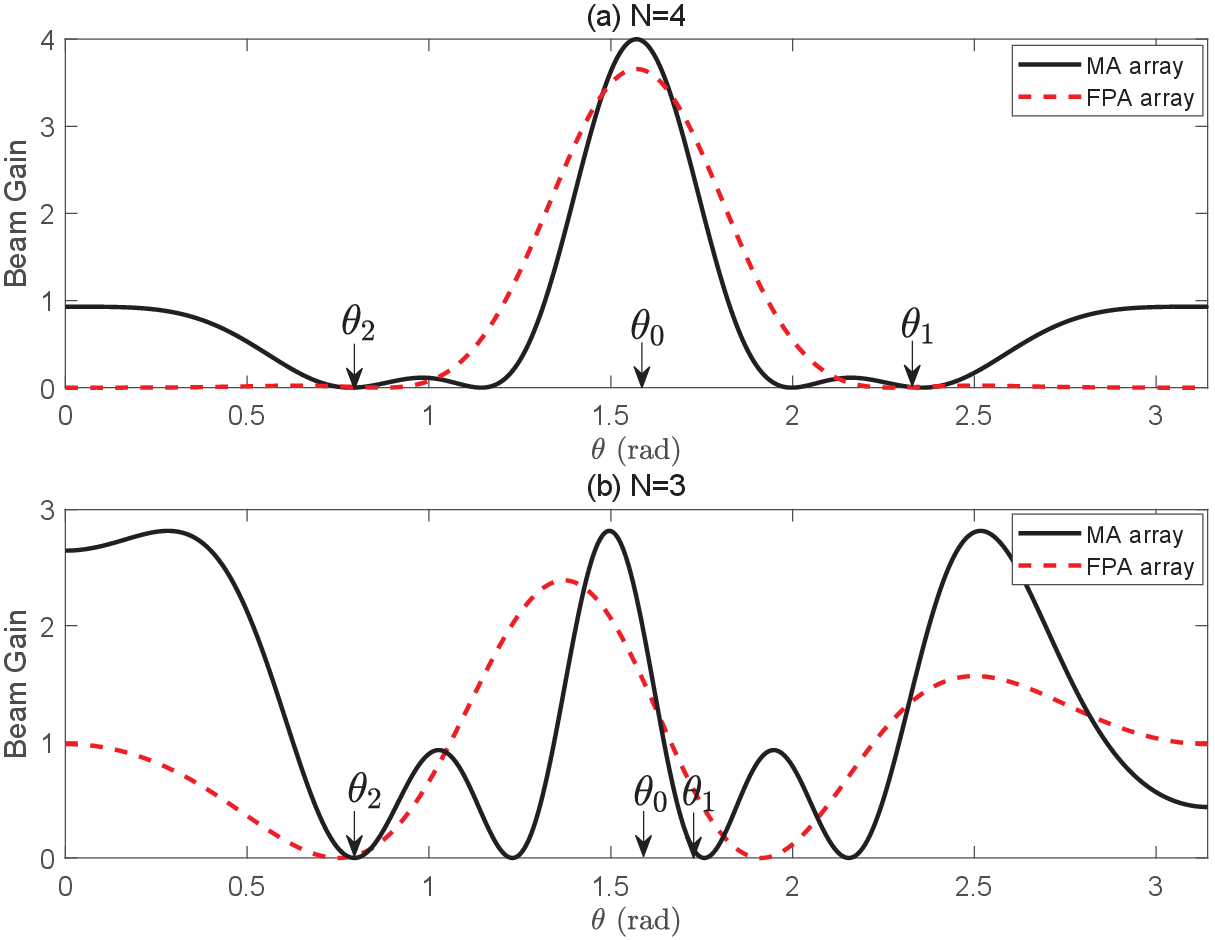}
\captionsetup{font=small}
\caption{Beam gain of the MA array and the FPA array at the angle $\theta  \in [0,\pi ]$.} \label{fig:Fig1}
\vspace{-10pt}
\end{figure}
\vspace{-10pt}
\section{Simulation Results}
In this section, numerical results are presented to demonstrate the effectiveness of the MA array for enhancing the secrecy performance. The minimum distance between any two MAs is set as ${d_{\min }} = \lambda /2$, and the range of the MA array is $[0,L] = [0,10\lambda ]$. The step size for the PGA method is set as $\delta  = 0.01$. The noise power is set as ${\sigma ^2} = 1$ for normalizing the large-scale channel fading power.

Fig. 2 illustrates the convergence behavior of (a): the proposed PGA method for optimizing ${\bf{x}}$ given ${\bf{w}} = \sqrt {{P_A}} \frac{{{\bf{a}}({{\bf{x}}^I},{\theta _0})}}{{\left\| {{\bf{a}}({{\bf{x}}^I},{\theta _0})} \right\|}}$ with $P_A = 1$ and ${{\bf{x}}^I}$ mentioned in Algorithm 1; (b) the alternating algorithm for iteratively optimizing ${\bf{x}}$ and ${\bf{w}}$. The parameters are $M = 2$, ${\theta _0} = \frac{\pi }{2}$, ${\theta _1} = \frac{{3\pi }}{4}$ and ${\theta _2} = \frac{\pi }{4}$ for the $N = 4$ case, and ${\theta _0} = \frac{\pi }{2}$, ${\theta _1} = \frac{{1.1\pi }}{2}$ and ${\theta _2} = \frac{\pi }{4}$ for the $N = 3$ case. It can be clearly observed that the objective in the inner (outer) layer converges to a constant without exceeding 50 (4) iterations for both cases of $N = 4$ and $N = 3$, indicating that the proposed algorithms are computationally efficient.

Fig. 3 presents the optimized positions of the MA array given $N = 4$ and $N = 3$. The system parameters are the same as in Fig. 2. It is observed that the positions of the MA array do not follow a simple uniform distribution as the FPA array with the fixed positions as ${{\bf{x}}^{{\rm{FPA}}}} = [0,{d_{\min }},...,(N - 1){d_{\min }}]{}^T$.

Fig. 4 accordingly provides the beam gain of the MA array and the FPA array at the angle $\theta $, with $\theta  \in [0,\pi ]$. Note that for the FPA array, only the transmit beamforming is optimized via equation (7). It is shown that when $N = 4$, the MA array and the FPA array concurrently achieve null steering at the undesirable angles (${{\theta _1}}$ and ${{\theta _2}}$) where two Eves are located. While by exploiting the additional DoF in flexible antenna positions, the MA array produces a higher gain at the desirable angle (${{\theta _0}}$) where Bob is located. On the other hand, when $N = 3$, the beamforming ability of both antenna arrays deteriorates. Still, the MA array almost achieves null steering at angles ${{\theta _1}}$ and ${{\theta _2}}$ and the preferable beam gain at the angle ${{\theta _0}}$, while the FPA array leakages the signal power at the angle ${{\theta _1}}$ and also achieves a smaller beam gain at the angle ${{\theta _0}}$.

Fig. 5 shows the achievable secrecy rate of the MA array and the FPA array w.r.t. number of antennas at Alice ($N$) under different power budget of Alice, where $M = 3$, ${\theta _0} = \frac{\pi }{2}$, ${\theta _1} = \frac{\pi }{4}$, ${\theta _2} = \frac{{0.85\pi }}{2}$ and ${\theta _3} = \frac{{1.1\pi }}{2}$. It is observed that as $N$ increases, since the larger spatial diversity and multiplexing gains are realized, obviously the secrecy rate of the MA array and the FPA array increases accordingly. In addition, armed with the flexible positions of all antennas, the MA array achieves a significant performance gain compared to the FPA array, indicating the great potential of the MA array for enhancing the communication security.

 \begin{figure}
 \vspace{-1pt}
\centering
\includegraphics[width=7cm]{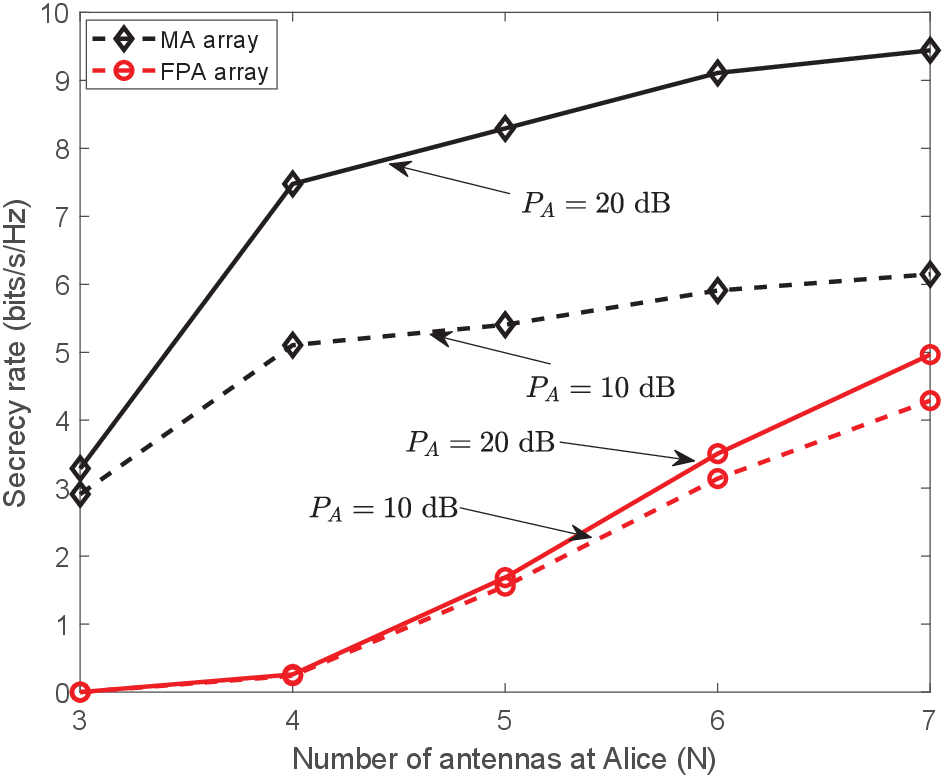}
\captionsetup{font=small}
\caption{Achievable secrecy rate w.r.t. number of antennas at Alice ($N$) based on the MA array and the FPA array.} \label{fig:Fig1}
\vspace{-20pt}
\end{figure}

\vspace{-10pt}
\section{Conclusion}
This letter studies a novel MA array-enabled secure communication system, and develops a PGA method and an alternating optimization algorithm to jointly optimize the transmit beamforming and the positions of all movable antennas at Alice for maximizing the achievable secrecy rate. It is shown by simulations that since the MA array brings additional DoF via antenna position optimizations, such array significantly enhances the security performance compared to the conventional FPA array.

\end{document}